\documentclass[fleqn,twoside]{article}
\usepackage{espcrc2}
\usepackage{graphicx}
\usepackage[figuresright]{rotating}
\usepackage{epsfig}

\newcommand{\numucc}{$\nu_{\mu}~CC$~}
\newcommand{\anumucc}{$\bar\nu_{\mu}~CC$~}

\newcommand{\bp}{$Bp$~}
\newcommand{\bpim}{$B\pi^{-}$~}
\newcommand{\bmath}{\begin{displaymath}}
\newcommand{\emath}{\end{displaymath}}
\newcommand{\be}{\begin{equation}}
\newcommand{\ee}{\end{equation}}

\title{A study of nuclear effects in $\nu$ interactions
       with the NOMAD detector}

\author{M. Veltri\address{Istituto di Fisica, Universit\`a di Urbino 
        and INFN Sezione di Firenze \\ 
        V. S. Chiara 27, I-61029 Urbino, Italy}%
  \thanks{On behalf of the NOMAD collaboration}}
       
\begin{document}

\begin{abstract}
 Nuclear effects in \numucc interactions with carbon nuclei
 have been studied by using backward going protons and $\pi^-.$ 
 Detailed analyses, of the momentum distributions and of the production
 rates, have been carried out in order to understand the mechanism
 producing these particles. The backward proton data have been compared
 with the predictions of the reinteraction and the short range
 correlation models.
\vspace{1pc}
\end{abstract}

\maketitle

\section{Introduction}
Since a long time \cite{baldin} it has been observed that 
in the high energy interactions off nuclei there are particles 
emitted backwards, with respect to the beam direction, which 
have energies not allowed by the kinematics of collisions on a 
free and stationary nucleon.
Backward going protons are commonly observed while, in absence of
nuclear effects, their production is forbidden. High energy mesons,
whose production in the backward direction is only allowed up to a given
momentum, are detected also, at momenta above such a limit.
Essentially two types of models have been proposed to
explain the origin of these particles.
The first is based on the intranuclear cascade (INC) mechanism.
The production of particles in the kinematically forbidden region
is seen as the result of multiple scattering and of interactions of
secondary hadrons, produced in the primary $\nu$-nucleon collision, 
with the other nucleons while they propagate through the nucleus
\cite{kopel,ferrari-ranft}.
The second approach explains the production of backward particles
as the result of collisions off structures with mass larger than 
the nucleon one. 
These structures are formed, at small interparticle distance,
 under the action of the short range part of
the nuclear force. They may be described as clusters of a few
correlated nucleons \cite{fs1} or multiquarks clusters \cite{carlson,multiq}. 
The nucleons in these structures can acquire high momenta and
the fast backward going particles can be seen as a direct manifestation
of the  high momentum tail of the Fermi distribution.\\
These two classes of models  are not mutually exclusive and
the production in  the backward hemisphere can have contributions 
from both mechanisms.\\

\section{The NOMAD detector and event selection}
The NOMAD detector \cite{nomdet} consisted of an active
target of 44 drift chambers, $3\times3 m^{2}$ each, of
low average density ($0.1~g/cm^{3}$) and overall mass of 2.7 tons.
The drift chamber composition was made by over 90\% of the total weight
by carbon and elements with nearby atomic number.
Downstream the beam the target was followed by a transition radiation
detector which provided electron identification, a preshower and an
electromagnetic calorimeter. All these systems were located inside a
dipole magnet which created a magnetic field of $0.4~T$.
A hadron calorimeter and two muon stations were located just behind
the magnet coil. NOMAD collected data from 1995 to 1998. During the run a
total of $5.1~\times~10^{19}$ protons on target, corresponding to about 
$1.3~\times~10^{6} \nu_{\mu}~CC~$ interactions, were recored on tape.
The events are selected  by requiring one primary vertex with at least
two tracks,
\begin{figure*}[ht]
\begin{center}
 \vskip -0.5cm
 \epsfig{file=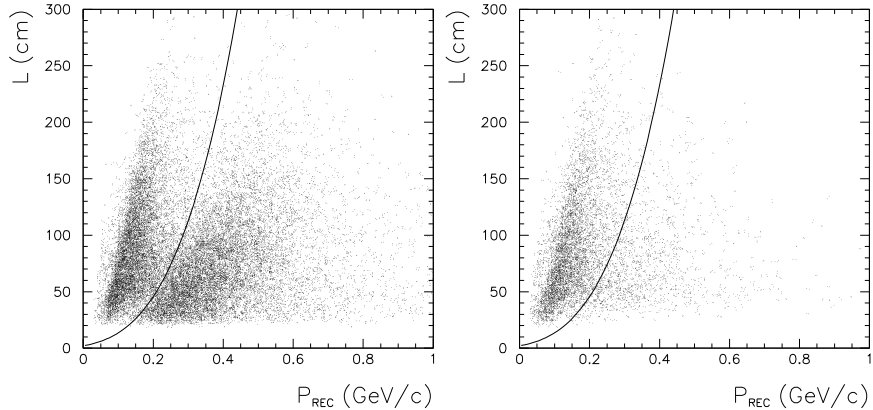,width=123mm}
\end{center}
\vskip -1.2cm
\caption{{\it Distributions of length vs. momentum for positive
(left) and negative (right) backward going tracks selected as described
in the text. The line indicates the position of the cut. }}
\label{lenvsp}
\end{figure*}
one of which has to be a negative muon of momentum greater 
than 3 GeV/c.
The sample selected for this study amounts to 944019 events.

\section{Backward particle identification}
\subsection{Backward proton identification}
The separation of the backward going protons ($Bp$) from the other
backward going positive tracks (essentially $\pi^{+}$)
is done by exploiting the range--momentum relation.
For this reason the candidate tracks are required to range out
in the drift chamber volume.
Fig.~\ref{lenvsp} shows the experimental distributions
of length vs. momentum for positive (left) and negative (right)
tracks, going backward with respect to the beam direction, and
satisfying our selection criteria.
Two distinct populations are clearly visible on the positive
sample plot. Protons, having a shorter range than $\pi^+$, tend to
accumulate in the lower right part of the plot while the $\pi^+$'s tend
to populate the left--hand side. Comparing the two plots we see that the
lower right part of the negative sample is much less populated than the
corresponding region of the positive one, these tracks being mainly
$\pi^-$. 
Any positive backward going track with length smaller than the cut
(the line on Fig.~\ref{lenvsp}) is identified as a $Bp$. The identified 
sample is then corrected for reconstruction efficiencies and for
$\pi^+$ contamination ($\approx 8\%$ for $P_{REC} > 250 MeV/c$).
For further details on the identification and correction procedure 
refer to \cite{bp_paper}.
\vskip -0.5cm

\subsection{Backward $\pi^{-}$ identification}
In the case of backward going $\pi^{-}$ ($B\pi^{-}$) it is not necessary
to look for tracks ranging out in the target. 
The choice of the negative charge allows for a sample of high purity;
it is shown by MC studies that the $e^{-}$ contamination is at the
level of few percent above a momentum of $\approx 200~MeV/c$.  
We therefore assume that any negatively, backward going,
charged track with $P_{REC} > 0.2~GeV/c$ is a $\pi^-$.
The identified sample is then corrected for reconstruction
efficiency.

\section{Backward $p$ and $\pi^-$ invariant momentum distributions}
\label{invslope}
The inclusive spectrum of backward particles is typically represented
using the normalized invariant cross section 
\begin{figure*}[t]
 \begin{center}
   \epsfig{file=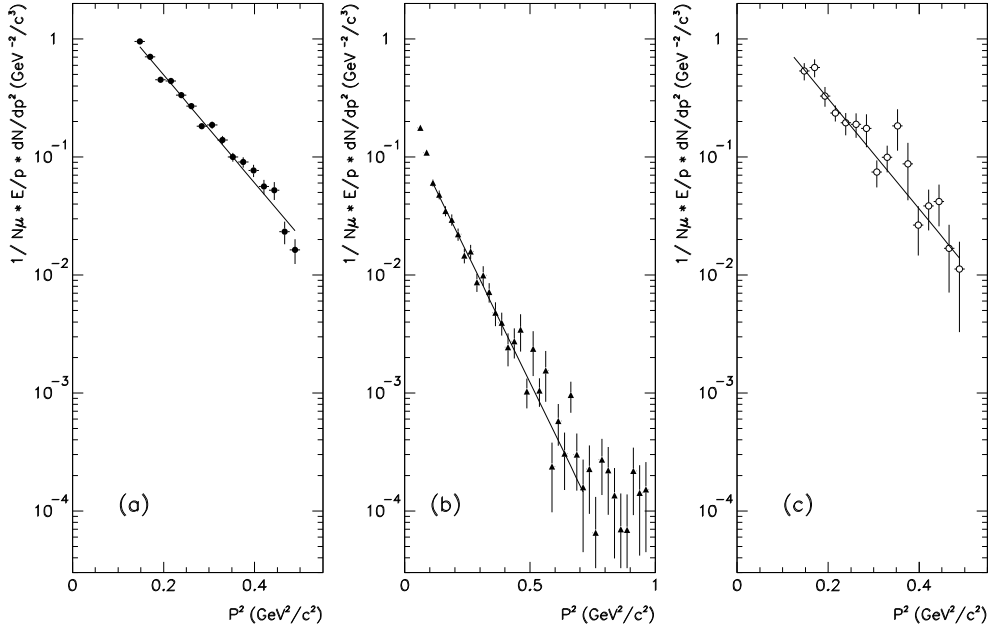,width=142mm}
 \end{center}
 \vskip -1.5cm
 \caption{{\it Invariant momentum distributions for backward going
    protons (a), $\pi^-$ (b) in $\nu_\mu CC$ events and protons in 
    $\bar\nu_\mu CC$ events (c)}}
 \label{slopes}
\end{figure*}
$(1/\sigma_{TOT})~E~d^3\sigma/dP^3$,
where $E$ is the energy of the backward going particle. The invariant
cross section is usually parametrized by an exponential form as:
\be
{1 \over N_{ev} }~{E \over P}~{dN \over dP^{2}}~=~C~e^{-BP^{2}}
\label{bparam}
\ee
\noindent where $N_{ev}$ is the total number of events,
$C$ is a constant and $B$ is the slope parameter. Early hadronic
experiments had found that $B$ is almost independent of the projectile 
type and momentum, and of the atomic number of the target.
This behaviour has been termed ``{\em nuclear scaling}''.
Neutrino experiments \cite{berge}-\cite{dayon} have confirmed 
these properties.\\
The inclusive spectrum of \bp and \bpim is shown in Fig.~\ref{slopes}a
and \ref{slopes}b respectively, together with the exponential fit.
In the \bpim case we have not included the first two points in the fit.
These points are in a momentum region where, backward production
being kinematically allowed, there are additional contributions
not coming from nuclear effects.
The values of the measured slope parameter $B$ are reported in
Table~\ref{table:slope_table}. The first and the second errors are
statistical and systematical, respectively.
The systematic uncertainty was estimated by slightly changing
the values of all the cuts used, by varying by a small amount the 
correction functions and also by changing the size of the fiducial volume. 
The values of the slope parameters measured in this experiment are found 
to be compatible with the results obtained in other neutrino 
and hadronic experiments.
\noindent The invariant cross section for \bp is larger than the one for
\bpim by about one order of magnitude but the values of the slopes are
similar. The kinematic ranges of the two invariant distributions are also 
different. To be identified \bp have to stop inside the target volume; the 
rather low density of the NOMAD target restricts to $\approx 0.7~GeV/c~$
the maximum useful momentum value.\\
\noindent The invariant cross section and slope parameter for \bp in 
\anumucc events are given in Fig.~\ref{slopes}c and 
Table~\ref{table:slope_table} respectively.
\begin{table*}[ht]
\caption{{\it Fit ranges and values of the slope parameter $B$, for backward protons 
and $\pi^{-}$, as obtained from the exponential fit to the invariant cross section.
The first and the second errors are statistical and systematical, respectively.}}
\label{table:slope_table}
\renewcommand{\tabcolsep}{1pc} 
\renewcommand{\arraystretch}{1.2} 
\begin{center}
\begin{tabular}{llcll}
\hline
                 &          & $\Delta P~(GeV/c)$ & $~~~B (c^{2}/GeV^{2})$ & $~~~C (c^3/GeV^2)$  \\ \hline
Backward $p$     & \numucc  & $0.37 \div 0.70$   & $10.54\pm0.20\pm0.5$ & $4.08\pm0.19\pm0.5$ \\
                 & \anumucc & $0.37 \div 0.70$   & $10.79\pm0.78$       & $2.71\pm0.54$       \\
Backward $\pi^-$ & \numucc  & $0.32 \div 0.85$   & $10.03\pm0.28\pm0.3$ & $0.17\pm0.01\pm0.02$\\
\hline \\
\end{tabular}
\end{center}
\end{table*}
During normal operations \anumucc events were also collected due to the small
$\bar\nu_{\mu}$ component of the dominantly $\nu_{\mu}$ beam.
A dedicated $\bar\nu$ run yielded an additional sample of \anumucc events
included for this analysis.\\
Antineutrino events are selected under the assumptions that the 
efficiencies and pion contamination are the same as those used for 
the neutrino events, but requiring a positively charged muon instead 
of a negative one. The final \anumucc sample consists of $61134$ events.

\section{Energy dependence of the slope parameter}
\begin{figure}[htb]
  \vskip -1.0cm
  \epsfig{file=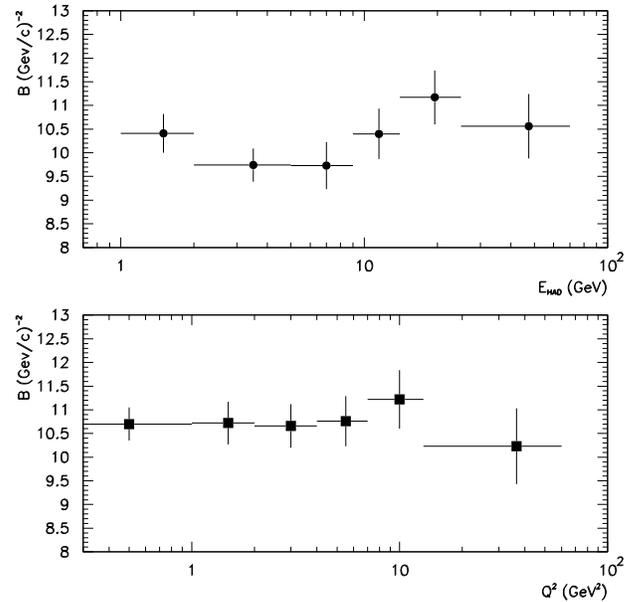,width=90mm}
  \vskip -1.0cm
  \caption{{\it The value of the slope parameter $B$ shown as a function
          of the hadronic energy $E_{HAD}$ (top) and of $Q^2$ (bottom). }}
  \label{behad}
\end{figure}

In Fig.~\ref{behad} we show the slope parameter $B$ for protons as a function
 of the visible hadronic energy $E_{HAD}$ and the $Q^2$ of the event.
The visible hadronic energy is defined as:
\bmath
E_{HAD}= \sum E_{c} + \sum E_{n}
\emath
\noindent where $\sum E_{c}$ is the sum of the energies of reconstructed 
charged tracks (assuming the mass of the pion if the particle is not 
otherwise identified) and $\sum E_{n}$ includes tracks associated to
secondary vertices corresponding to interactions of neutral particles and
the energy of neutral particles reconstructed in the ECAL. 
$Q^2$ is the square of the four--momentum transfer,
$Q^2 = 4~E_{vis}E_{\mu}sin^2\theta/2$, where $\theta$ is the muon angle
with respect to the neutrino direction.
\noindent No significant dependence of the slope $B$ on either quantity is 
observed, in agreement with the expectations of ``nuclear scaling'' as 
observed in hadronic beam experiments. The range covered by NOMAD is similar 
to the one covered by different experiments with hadronic beams.

\section{Backward particle rates}
\begin{figure}[ht]
\begin{center}
   \epsfig{file=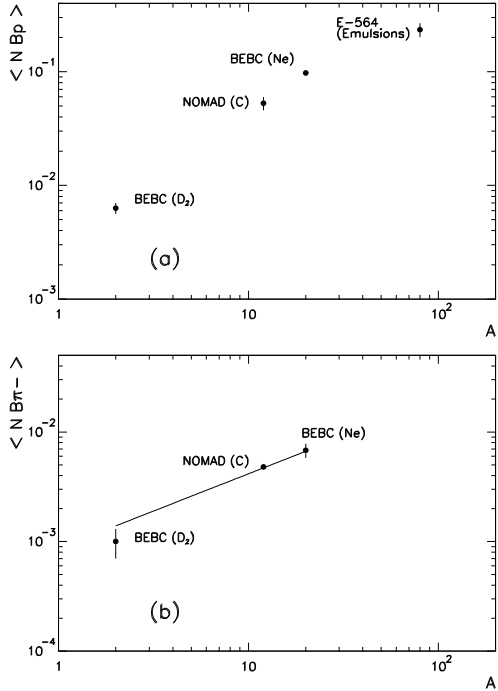,width=75mm}
\end{center}
\vskip -1.0cm
\caption{{\it The average number of \bp per event (a) and of \bpim 
per event (b) in the momentum range from  $350\div800 MeV/c$
in neutrino experiments as a  function of the atomic number $A$.
The line shown in (b) is the result of the fit described in the text. }}
\label{bpyield}
\end{figure}

In order to study a possible atomic number dependence
the \bp rate has been compared with the results of the other 
$\nu$--nucleus experiments.
The measured yields have all been normalized to the
same momentum interval (350 to 800 MeV/c) assuming the dependence given 
in  Eq.~\ref{bparam} with the measured slopes.\\
The extrapolation for NOMAD gives:
\bmath
<N_{Bp}>_{350\div800 MeV/c}~= \nonumber 
\emath
\bmath
\left[ \rule{0mm}{3mm}~52.8 \pm 0.6 (stat.) \pm 7 (syst.)~ \right] \times 10^{-3}
\nonumber
\emath
The $A$ dependence for experiments most directly comparable to NOMAD
is shown in Fig.~\ref{bpyield}a. In the range $A=20\div80$
it has been parametrized as $ <N_{BP}>\propto A^{\alpha}$,
where $\alpha=0.68\pm0.12$ \cite{dayon}.
The same parametrization with a similar power law was found to describe
\bp production in $\pi^{+}$ and $K^{+}$ collisions with $Al$ and $Au$
nuclei at 250 GeV/c \cite{na22}.
It is evident from Fig.~\ref{bpyield}a that this simple power law does not
describe the NOMAD data taken at a lower $<A>$.
The \bpim rate was directly measured in the same momentum range.
Its average value is found to be:
\begin{displaymath}
<N_{B\pi^-}>_{350\div800 MeV/c}~ =
\end{displaymath}
\begin{displaymath}
 \left[ \rule{0mm}{3mm}  ~4.8 \pm 0.1 (stat.) \pm 0.3 (syst.) \right]
\times 10^{-3}
\end{displaymath}
\noindent In this case a good fit to the two BEBC points and to the NOMAD value
can be obtained using the form $A^{\alpha}$ giving $\alpha=0.83\pm0.25$,
as shown in Fig.~\ref{bpyield}b.

\section{Comparison of data with models}
\label{data&model}
\subsection{$E_{HAD}$ and $Q^2$ dependence of \bp and \bpim rates}
\begin{figure*}[t]
\vskip -0.5 cm
\begin{center}
   \epsfig{file=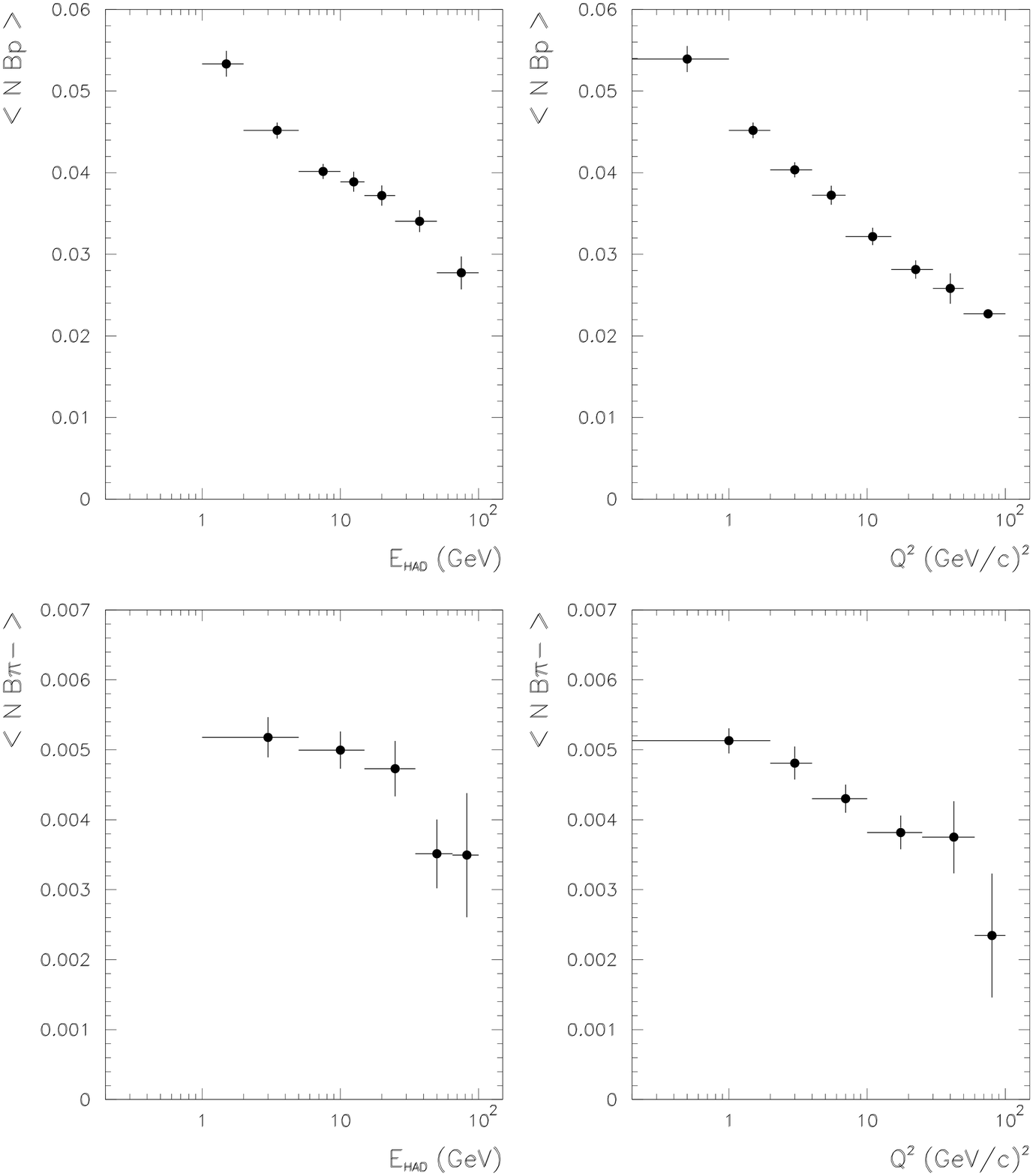,width=100mm}
\end{center}
\vskip -1.0cm
\caption{{\it Top picture: the average number of \bp ($ 370<P<700 MeV/c $)
per event as a function of the hadronic energy $E_{HAD}$ (left) and of $Q^2$ (right).
Bottom picture: the average number of \bpim ($ 350<P<800 MeV/c $)
per event as a function of the hadronic energy $E_{HAD}$ (left) and of $Q^2$ (right).}}
\label{bpy_ehad}
\end{figure*}

The \bp and the \bpim rates have been studied as a function of the hadronic energy
$E_{HAD}$ and of $Q^2$. In both cases, shown in Fig.~\ref{bpy_ehad},
a decrease of the yield with increasing $E_{HAD}$ and $Q^2$ is observed. 
This can be interpreted in terms of the ``formation zone'' concept. This is the
distance (or the time) from the production point which is required for
the secondary hadrons to be ``formed'', i.e. to be able to interact
as physical hadronic states. Since the distance/time required,
due to the Lorentz time--dilation factor,
is proportional to the energy of the secondary,
the INC process is restricted to slow hadrons which have formation
lengths smaller than the nuclear radius.
The larger $E_{HAD}$ and $Q^2$, the larger is the average energy of the outgoing
partons therefore resulting in hadrons which have higher probability to
be formed outside the nucleus.
As a consequence, reinteractions will decrease and so will the slow proton rates.
In the \bpim case the dependence of the yield on $Q^2$ and on $E_{HAD}$ is
less pronounced. This can be understood since the \bpim rates are a less
sensitive probe of nuclear effects because \bpim production on a stationary
nucleon is kinematically allowed for momenta up to about half the nucleon mass.
Furthermore \bpim can be produced in the decay of forward going resonances
or unstable particles.

\subsection{Effects of short range correlations in \bp production}

\begin{figure}[htb]
\vskip -1.0cm
\begin{center}
   \epsfig{file=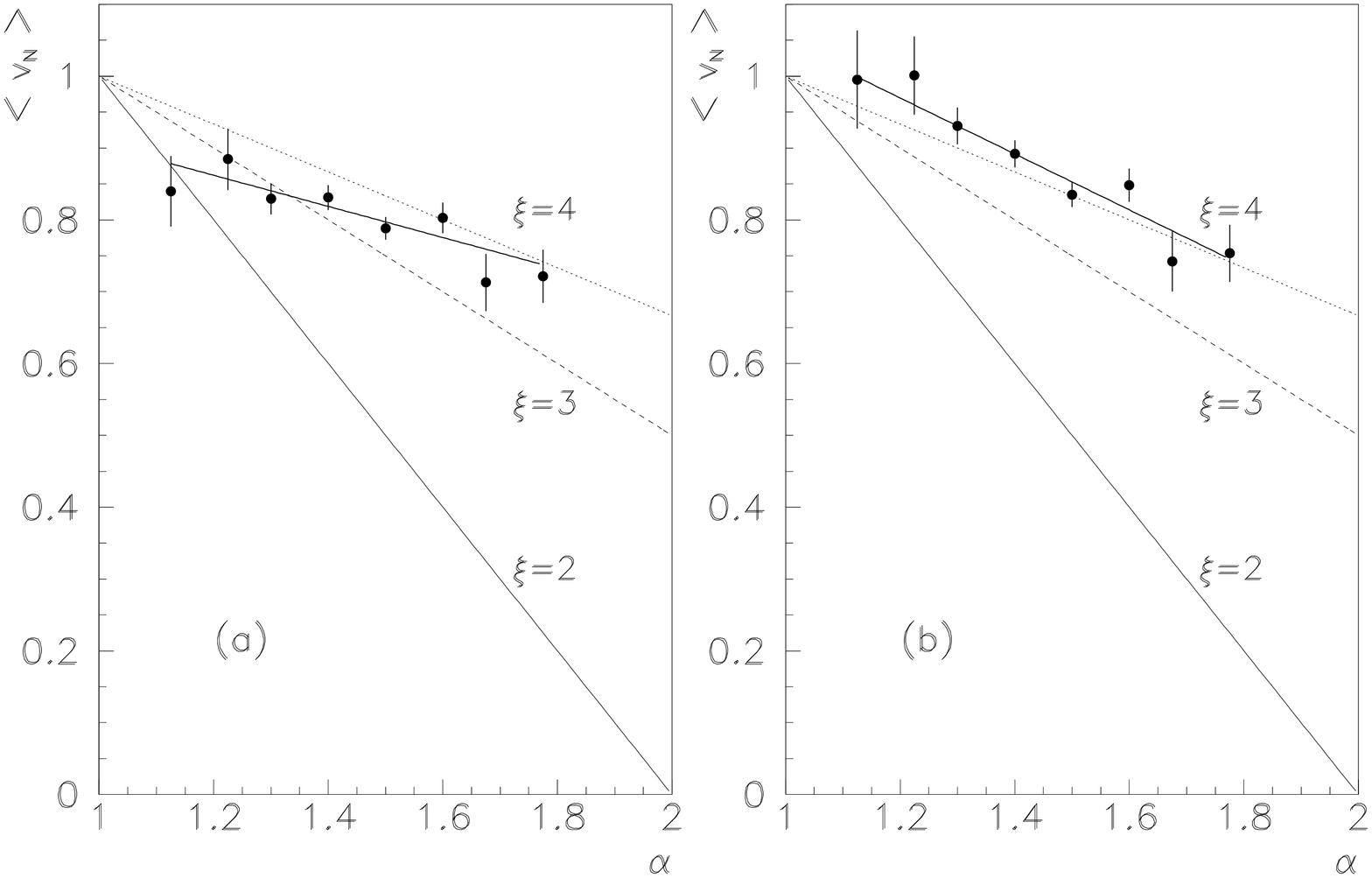,width=85mm}
\end{center}
\vskip -1.cm
\caption{{\it The variable $<v_N>$ plotted as a function of $\alpha$.
The lines represent the predicted correlation (Eq.~\ref{v_cluster})
for a number $\xi$ of nucleons in the cluster equal to 2, 3 and 4.
In (a) all the \bp events were used;
in (b) only events having a \bp with $cos{\theta_{j}} < 0$, $\theta_{j}$
being the angle of the \bp with respect to the hadronic jet direction.}}
\vskip -0.5 cm
\label{alpha}
\end{figure}

According to the picture proposed by the correlated nucleon/quark
models the \bp production is explained as the result of a neutrino
interaction within a correlated cluster of nucleons/quarks.
This cluster is formed, for a short time, when two nucleons in their motion
inside the nucleus approach each other so closely as to come under the effect
of the short range component of the nuclear force ($r_c = 0.5 \div 0.7 fm$).
The repulsive character of this interaction is responsible of the high
relative momenta of the cluster members.
The simplest, and the most probable case when considering the 
overlapping probabilities of nucleons in the nucleus, is the two--nucleon
case. According to this model the \bp is released when the correlated pair
is broken in the $\nu$--cluster interaction.
If the effects of reinteractions are neglected, the released \bp 
can leave the nucleus keeping its original momentum.
To study these correlations it is customary to use the variable
$\alpha$ defined as:
\be
\alpha=(E~-P_l)/ M
\label{eq_alpha}
\ee
 \noindent where $E$ and $P_l$ are respectively
the energy and  longitudinal momentum of the \bp and
$M$ is the nucleon mass. For $Bp$, $\alpha > 1$ since $P_l$ is negative.
In this model, due to the target motion a correlation between
$\alpha$ and the variable $v = xy$, where $x$ is the Bjorken scaling
variable $x=Q^{2}/2ME_{HAD}$ and $y=E_{HAD}/(E_{HAD}+E_{\mu})$ 
is expected. The variable $v$ is related to the muon kinematics 
and can be written as:
\be
v=(E_{\mu}-P_\mu^{l})/M
\label{eq_v}
\ee
\noindent where $E_{\mu}$ and $P_\mu^{l}$ are the muon energy and longitudinal
momentum. The $(v,\alpha)$ correlations were searched for in the data
by calculating for each event $\alpha$ and $v$ as defined in Eq.~\ref{eq_alpha} and
\ref{eq_v}. For each $\alpha$ bin we plot the variable $<v_N>$ defined as:
\be
<v_N>~=~{<v>_{Bp} \over <v>_{no~Bp}}
\ee
\noindent where $<v>_{no~Bp}$ is obtained from the full sample of events
without a $Bp$. According to Ref.~\cite{fs1} the average values of
$v$ in events with a $Bp$, $<v>_{Bp}$, is related to the average value of $v$ in events
where no \bp is present, $<v>_{no~Bp}$, by:
\be
<v>_{Bp}~=~<v>_{no~Bp}~(2-\alpha)
\ee
\noindent More generally for a cluster composed of $\xi$ nucleons
the relation is \cite{multiq}:

\be
<v>_{Bp}~=~<v>_{no~Bp} \left(1 - {\alpha \over \xi} \right)~ {\xi \over {\xi-1}}
\label{v_cluster}
\ee
\noindent Fig.~\ref{alpha}a shows  $<v_N>$ as a function of $\alpha$.
The data indicate a slope of $-0.22\pm0.06$ with a $\chi^{2}/ndf=5.3/6$
(the $\chi^{2}/ndf$ in the hypothesis of no dependence on $\alpha$ is 19.1/7).
The two--nucleon correlation mechanism
fails to describe our data. Either higher order structures are playing
a leading role \cite{multiq} or the observed low level of correlation is due
to the presence of reinteraction processes. 
\noindent  In this case part of the
\bp are emitted as a result of reinteractions in the nucleus and are not
related to the target nucleon. The presence of intranuclear cascade processes
could therefore dilute the existing correlation to the observed level.
To test this hypothesis we tried to reduce the component
due to rescattering in the selected \bp sample. Having observed the
correlation existing between the multiplicities of slow tracks
and rescattered protons we applied increasingly
tighter cuts on the number of slow tracks ($P<700~MeV/c$).
As a consequence of these cuts the degree of correlation between
$\alpha$ and $v$ increases.
The fit values of the slopes are reported in Table~\ref{table:alpha}
together with the definitions of the cuts applied.
We have also observed a strong correlation between the presence of 
protons travelling backward in the lab but forward with respect to the
hadronic jet direction, and the concentration of events  at small $Q^2$
values and large angles with respect to the beam.
Since also a small $Q^2$ indicates the presence of rescattering,
the exclusion of these events should highlight the expected correlation.
The resulting slope is $-0.39\pm0.07$ with a $\chi^{2}/ndf=4.9/6$
(see Fig.~\ref{alpha}b).
The observed behaviour is consistent with the hypothesis of the correlations
effects being to some degree hidden by the presence of rescattering.
Reducing the rescattering component these correlations seem to become stronger.
\vskip -.2 cm
\begin{table}[ht]
\caption{\it { The fitted value of the ($\alpha, v$) slope,
and the corresponding $\chi^{2}/ndf$, for \bp
selected from events with various numbers of positive ($n^+$)
and negative ($n^-$) low  momentum ($P < 700 MeV/c$) particles.}}
\label{table:alpha}
\centering
\begin{tabular}{cclc}
\hline
 ~~$n^+$   & ~~$n^-$    & ~~~~~~~slope & $\chi^{2}/ndf$ \\ \hline \smallskip
 $\leq~3$  & $\leq~2$ & $-0.23\pm0.06$ & 5.7/6  \\
 $\leq~2$  & $\leq~1$ & $-0.25\pm0.07$ & 3.3/6 \\
 $\leq~2$  &       0  & $-0.30\pm0.07$ & 3.7/6 \\
       1   &       0  & $-0.37\pm0.10$ & 3.9/6 \\ \hline
\end{tabular}
\vskip -0.6 cm
\end{table}
\section{Comparison with the NOMAD Monte Carlo}

The NOMAD event generator NEG--N is based on a modified version of 
LEPTO 6.1 \cite{lepto} and JETSET 7.4 \cite{jetset} (with fragmentation
parameters tuned on the NOMAD data) for the deep inelastic simulation 
and on dedicated generators for quasielastic events and events in 
the resonance region.
Nuclear effects are taken into account by incorporating the 
\begin{figure}[h]
\vskip -0.7 cm
\begin{center}
   \epsfig{file=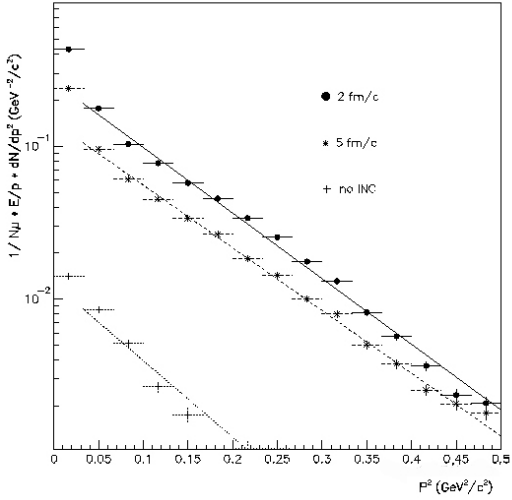,width=80mm}
\end{center}
\vskip -4.5cm
\caption{{\it Invariant momentum distributions of \bp simulated
with NEG--N for different values of the formation time parameter.}}
\vskip -0.8cm
\label{slopeda}
\end{figure}
formation zone intranuclear cascade (FZIC) code of DPMJET--II.4
\cite{dpmjet}. For the hadrons produced
inside the nucleus the code performs a complete simulation of tracking
and of interactions in nuclear matter developing the cascade
through several generations. The Fermi momentum is simulated according 
to \cite{benhar}.
The event generator used for the oscillation search did not take
into account the INC. 
In that case a discrepancy of $200 MeV/c$ on the average transverse
missing momentum, $p_{T}^{m}$ (the difference
between the reconstructed hadron jet and muon transverse momenta), 
was found between the data and the MC prediction.
In view of its importance in the $\tau$ search, based on the
exploitation of the different kinematics of the $\nu_{\tau}$ interactions
with respect to the $\nu_{e}$ and $\nu_{\mu}$ ones in the transverse
plane, the MC sample could not be directly used to calculate background
and signal efficiencies. However, a method \cite{datasimul} was
succesfully developed in order to correct the MC predictions using the
data themselves. 
The results of this upgraded generator show how the INC  
degradates the hadronic energy by producing $n$ and slow $p$
(emitted at large  angles and difficult to detect). The simulation of
nuclear effects reduces by a factor of two the  $p_{T}^{m}$ discrepancy
between data and MC; the remaining effect is probably due to a too
optimistic detector simulation. In the code the hadron formation
length is left as a free parameter to be supplied by the user. 
It has been tuned on the NOMAD data in
order to achieve the best agreement between the simulated and the
experimental hadronic distributions of multiplicities, momentum spectra
and angular  distributions. The best agreement is found for a formation
time of 2 fm/c. For this value the MC predicts a rate of 
$<~N_{Bp}>_{350\div800 MeV/c}~= 53.0 \times 10^{-3}$ with a slope of 10
in excellent agreement with the experimental results. It was found also
(see Fig.~\ref{slopeda}) that the rate is strongly dependent on the 
formation time while the slope is not.



\begin{thebibliography}{99}
\bibitem{baldin} A. Baldin et al., Sov. J. Nucl. Phys. {\bf 18} (1973) 41.
\bibitem{kopel}  V. B. Kopeliovich, Phys. Rep. {\bf 139} (1986) 51.
\bibitem{ferrari-ranft} A. Ferrari et al., Z. Phys. {\bf C 70} (1996) 413.
\bibitem{fs1} L. L. Frankfurt and M. Strikmann, 
Phys. Lett {\bf B 69} (1977) 93;  Phys. Rep. {\bf 76} (1981) 215.
\bibitem{carlson} C. E. Carlson, K. E. Lassila and U. P. Sukhatme, 
Phys. Lett. {\bf B 263} (1991) 277.
\bibitem{multiq} \noindent L. A. Kondratyuk and M. Zh. Shmatikov, 
Z. Phys. {\bf A 321} (1985) 301.
\bibitem{nomdet} J. Altegoer et al. NOMAD Coll., 
Nucl. Instr. and Meth. {\bf A 404} (1998) 96.
\bibitem{bp_paper} P. Astier et al. NOMAD Coll., 
Nucl. Phys. {\bf B 609} (2001) 255.
\bibitem{berge}
J. P. Berge et al., Phys. Rev. {\bf D 18} (1978) 1367;
V. I. Efremenko et al. Phys. Rev. {\bf D 22} (1980) 2581.
\bibitem{skat}
A. A. Ivanilov et al., JETP Lett. {\bf 30} (1979) 362.
\bibitem{ammosov}
V. V. Ammosov et al., Sov. J. Nucl. Phys. {\bf 43} (1986) 759.
\bibitem{matsinos}
E. Matsinos et al. Z. Phys. {\bf C 44} (1989) 79.
\bibitem{dayon}
M. Dayon et al. Z. Phys. {\bf C 56} (1992) 391.
\bibitem{na22}
N. M. Agababyan et al, EHS/NA22 Coll., Z. Phys. {\bf C 66} (1995) 385.
\bibitem{lepto}
G.~Ingelman, LEPTO 6.1, in Proc. of Physics at HERA,
DESY, Hamburg (1992) 1366.
\bibitem{jetset}
T.~Sj\"ostrand, Computer Phys. Commun. {\bf 39} (1986) 347.
\bibitem{dpmjet}
J. Ranft, Gran Sasso Lab. Report, INFN/AE--97/45 (1997).
\bibitem{benhar}
O. Benhar, private communication.
\bibitem{datasimul}
J. Altegoer et al., NOMAD Coll., Phys. Lett {\bf B 431} (1998) 219.


\end{thebibliography}
\end{document}